\documentclass[preprint,12pt]{elsarticle}

\usepackage{amssymb}
\usepackage{multirow}
\usepackage{color}

\journal{Nuclear Physics A}

\newcommand{\PerDay}{day$^{-1}$}
\newcommand{\PerTonDay}{(ton$\cdot$day)$^{-1}$}
\newcommand{\xe}{$^{136}$Xe}
\newcommand{\znu}{0$\nu\beta\beta$}
\newcommand{\tnu}{2$\nu\beta\beta$}

\begin{document}

\begin{frontmatter}



\title{Search for double-beta decay of {\xe} to excited states of $^{136}$Ba with the KamLAND-Zen experiment}


%
%

\author[tohoku]{K.~Asakura}
\author[tohoku]{A.~Gando\corref{cor}}
\ead{azusa@awa.tohoku.ac.jp}
\author[tohoku]{Y.~Gando}
\author[tohoku]{T.~Hachiya}
\author[tohoku]{S.~Hayashida}
\author[tohoku]{H.~Ikeda}
\author[tohoku,ipmu]{K.~Inoue}
\author[tohoku]{K.~Ishidoshiro}
\author[tohoku]{T.~Ishikawa}
\author[tohoku]{S. Ishio}
\author[tohoku,ipmu]{M.~Koga}
\author[tohoku]{S.~Matsuda}
\author[tohoku]{T.~Mitsui}
\author[tohoku]{D.~Motoki}
\author[tohoku,ipmu]{K.~Nakamura}
\author[tohoku]{ S.~Obara}
\author[tohoku]{M.~Otani\fnref{kek}}
\author[tohoku]{T.~Oura}
\author[tohoku]{I.~Shimizu}
\author[tohoku]{Y.~Shirahata}
\author[tohoku]{J.~Shirai}
\author[tohoku]{A.~Suzuki}
\author[tohoku]{H.~Tachibana}
\author[tohoku]{K.~Tamae}
\author[tohoku]{K.~Ueshima}
\author[tohoku]{H.~Watanabe}
\author[tohoku]{B.D.~Xu\fnref{ipmu1}}
\author[tohoku]{H.~Yoshida\fnref{osakau}}

\cortext[cor]{Corresponding author. }
\fntext[kek]{Present address: High Energy Accelerator Research Organization (KEK), Tsukuba, Ibaraki, Japan}
\fntext[ipmu1]{Present address: Kavli Institute for the Physics and Mathematics of the Universe (WPI), The University of Tokyo Institutes for Advanced Study, The University of Tokyo, Kashiwa, Chiba 277-8583, Japan}
\fntext[osakau]{Present address: Graduate School of Science, Osaka University, Toyonaka, Osaka 560-0043, Japan}

\author[ipmu]{A.~Kozlov}
\author[ipmu]{Y.~Takemoto}

\author[osaka]{S.~Yoshida}

\author[tokushima]{K.~Fushimi}

\author[lbl]{T.I.~Banks}
\author[lbl,ipmu]{B.E.~Berger}
\author[lbl,ipmu]{B.K.~Fujikawa}
\author[lbl]{T.~O'Donnell}

\author[mit]{L.A. Winslow}

\author[ut,nrnu,ipmu]{Y.~Efremenko}

\author[tunl]{H.J.~Karwowski}
\author[tunl]{D.M.~Markoff}
\author[tunl,ipmu]{W.~Tornow}

\author[washington,ipmu]{J.A.~Detwiler}
\author[washington,ipmu]{S.~Enomoto}

\author[nikhef,ipmu]{M.P.~Decowski}


\address[tohoku]{Research Center for Neutrino Science, Tohoku University, Sendai 980-8578, Japan}
\address[ipmu]{Kavli Institute for the Physics and Mathematics of the Universe (WPI), The University of Tokyo Institutes for Advanced Study, The University of Tokyo, Kashiwa, Chiba 277-8583, Japan}
\address[osaka]{Graduate School of Science, Osaka University, Toyonaka, Osaka 560-0043, Japan}
\address[tokushima]{Faculty of Integrated Arts and Science, University of Tokushima, Tokushima, 770-8502, Japan}
\address[lbl]{Physics Department, University of California, Berkeley, and \\ Lawrence Berkeley National Laboratory, Berkeley, California 94720, USA}
\address[mit]{Massachusetts Institute of Technology, Cambridge, Massachusetts 02139, USA}
\address[ut]{Department of Physics and Astronomy, University of Tennessee, Knoxville, Tennessee 37996, USA}
\address[nrnu]{National Research Nuclear University, Moscow, Russia}
\address[tunl]{Triangle Universities Nuclear Laboratory, Durham, North Carolina 27708, USA and Physics Departments at Duke University, North Carolina Central University, and the University of North Carolina at Chapel Hill}
\address[washington]{Center for Experimental Nuclear Physics and Astrophysics, University of Washington, Seattle, Washington 98195, USA}
\address[nikhef]{Nikhef and the University of Amsterdam, Science Park, Amsterdam, the Netherlands}

\begin{abstract}
A search for double-beta decays of {\xe} to excited states of $^{136}$Ba has been performed with the first phase data set of the \mbox{KamLAND-Zen} experiment.
The $0^+_1$, $2^+_1$ and $2^+_2$ transitions of {\znu} decay were evaluated in an exposure of 89.5\,kg$\cdot$yr of {\xe}, while the same transitions of {\tnu} decay were evaluated in an exposure of 61.8\,kg$\cdot$yr. No excess over background was found for all decay modes. 
The lower half-life limits of the $2^+_1$ state transitions of {\znu} and {\tnu} decay were improved to $T_{1/2}^{0\nu}(0^+ \rightarrow 2^+_1) > 2.6\times10^{25}$ yr and $T_{1/2}^{2\nu}(0^+ \rightarrow 2^+_1) > 4.6\times10^{23}$ yr (90\% C.L.), respectively. 
We report on the first experimental lower half-life limits for the transitions to the $0^+_1$ state of {\xe} for {\znu} and {\tnu} decay. They are $T_{1/2}^{0\nu}(0^+ \rightarrow 0^+_1) > 2.4\times10^{25}$ yr and $T_{1/2}^{2\nu}(0^+ \rightarrow 0^+_1) > 8.3\times10^{23}$ yr (90\% C.L.). 
The transitions to the $2^+_2$ states are also evaluated for the first time to be $T_{1/2}^{0\nu}(0^+ \rightarrow 2^+_2) > 2.6\times10^{25}$ yr and $T_{1/2}^{2\nu}(0^+ \rightarrow 2^+_2) > 9.0\times10^{23}$ yr (90\% C.L.). These results are compared to recent theoretical predictions.
\end{abstract}


\end{frontmatter}


\section{Introduction}
\label{sec:Intro}
Double-beta decay is a rare nuclear process with fundamental connections to particle physics, nuclear physics and possibly cosmology. 
Double-beta decay can be a leading process in the decay of a nucleus if the first order decay is energetically forbidden or highly suppressed. Two neutrino double-beta ({\tnu}) decay, (A,Z) $\rightarrow$ (A, Z+2) + 2e$^-$ + 2$\bar{\nu_{\rm e}}$, is an established second order process in the standard model of the electroweak interaction. {\tnu} decay to the ground state has been observed in eleven nuclei with half-lives in the range from $10^{19}$ to $10^{24}$ yr~\cite{Barabash:2010ie,Gando:2012pj,Auger:2012ar}. The {\tnu} transition to the 0$^{+}_1$ excited state of the daughter nucleus was also measured for two nuclei ($^{100}$Mo and $^{150}$Nd)~\cite{Barabash:2010ie,PhysRevC.90.055501}, although it is strongly suppressed in comparison with the transition to the ground state due to the smaller transition energies. The measured half-life of {\tnu} decay can experimentally constrain the relevant nuclear matrix element (NME).
Measurement of different transitions provides useful information for the development of theoretical schemes to calculate NMEs. The understanding of {\tnu} decay NMEs is important for investigations of nuclear structure and of nuclear interactions associated with neutrinoless double-beta ({\znu}) decay~\cite{0954-3899-39-12-124005}.

Neutrino oscillation experiments demonstrated that neutrinos have non-zero mass and therefore the question arises whether neutrinos are Majorana or Dirac particles.
{\znu} decay is possible if the neutrino is a massive Majorana particle~\cite{Schechter:1981bd}.
Various experimental efforts exist to search for this lepton number violating process since it implies physics beyond the standard model.
In addition, if the decay is mediated by light Majorana neutrino exchange, the {\znu} decay rate can translate into an effective neutrino mass, $\left<m_{\beta\beta}\right> \equiv \left| \Sigma_{i} U_{ei}^{2}m_{\nu_{i}} \right|$, via the NME and a phase-space factor, where $U_{ei}$ is the corresponding element of the Pontecorvo-Maki-Nakagawa-Sakata mixing matrix. 
It would  constrain the absolute mass scale, and could provide information on the neutrino mass ordering.
To obtain the effective neutrino mass, detailed knowledge of the NME for {\znu} decay is essential.
The decay can proceed to the ground state as well as through transitions to excited states. Studies of these transitions provide additional information on the double-beta decay and associated  
NME calculations. 
Once {\znu} decay is observed, the measurement of the branching ratios of the {\znu} decay to the ground and excited states allow the possibility to distinguish between different {\znu} mechanisms, such as those mediated through light or heavy Majorana neutrinos, or the R-parity breaking mechanism~\cite{Simkovic:2001ft}.

\begin{figure}[htbp]
	\begin{center}
	\vspace{-1.0cm}
	\includegraphics[width=0.9\columnwidth,clip]{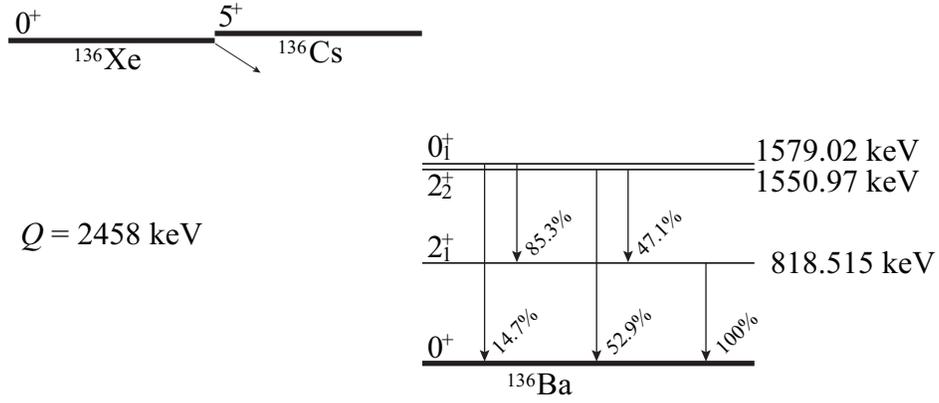}
	\vspace{-0.4cm}
	\end{center}
	\caption[]{Decay scheme of  {\xe} to $^{136}$Ba. The energy levels and the branching ratios for the de-excitations are taken from Ref.~\cite{TableOfIsotopes1999}. 
	Only the investigated 0$^+$ and 2$^+$ levels are shown. In the 0$^+_1 \rightarrow 0^+$ transition, conversion electron emission is dominant.
	}
	\label{figure:decayscheme}
	\end{figure}

In this article, we report limits on the double-beta decay of {\xe} to the 0$^{+}_1$, 2$^{+}_1$ and 2$^{+}_2$ excited states of $^{136}$Ba for both the {\tnu} and {\znu} decay modes. These limits were obtained with the first phase data set of the \mbox{KamLAND-Zen} (KamLAND ZEro-Neutrino double-beta decay) experiment. The decay scheme of {\xe} to $^{136}$Ba is shown in Fig.~\ref{figure:decayscheme}. The $Q$ value of {\xe} double-beta decay is 2458 keV~\cite{Redshaw:2007un}.

\section{The KamLAND-Zen experiment}
\label{sec:detector}
\begin{figure}[htbp]
	\begin{center}
	\includegraphics[width=0.9\columnwidth]{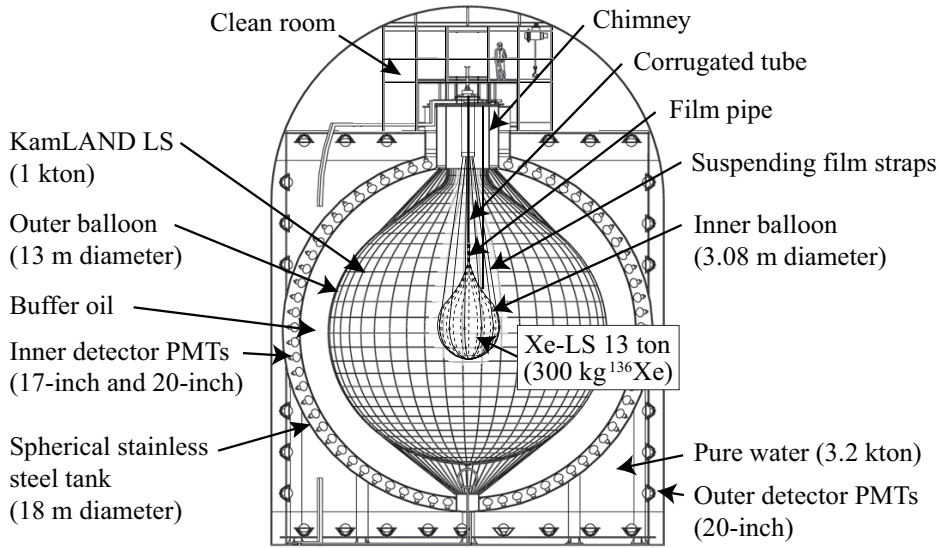}
	\vspace{-0.4cm}
	\end{center}
	\caption[]{Schematic diagram of the KamLAND-Zen detector.}
	\label{figure:detector}
	\end{figure}

The \mbox{KamLAND-Zen} experiment is a modification of the existing neutrino detector \mbox{KamLAND}~\cite{Abe:2009aa} located in the Kamioka mine, Gifu, Japan. 
A schematic diagram of the detector is shown in Fig.~\ref{figure:detector}. 
The initial phase of the \mbox{KamLAND-Zen} experiment uses 13\,tons of Xe-loaded liquid scintillator (Xe-LS) contained in a 3.08-m-diameter spherical inner balloon (IB), suspended at the center of the detector, as the double-beta decay source and detection medium. The IB is constructed from heat-welded 25-$\mu$m-thick transparent clean nylon film. 
Outside of the IB is 1\,kton of liquid scintillator (LS), which acts as an active shield for external $\gamma$'s and as a detector for internal radiation from the Xe-LS and/or IB. 
The LS is contained in a 13-m-diameter spherical outer balloon, made of 135-$\mu$m-thick nylon/EVOH\footnote[1]{Ethylene vinyl alcohol copolymer} composite film.
The density difference between Xe-LS and \mbox{KamLAND} LS is carefully controlled within 0.04\% to reduce the load on the IB. The Xe-LS consists of 82\% of decane and 18\% pseudocumene (1,2,4-trimethylbenzene) by volume, 2.7\,g/liter of the fluor PPO (2,5-diphenyloxazole), and 2.44 or 2.48\% by weight of enriched xenon gas for data set 1 and 2, respectively (see Section~\ref{sec:event_selection}).
The isotopic abundances in the enriched xenon were measured by a residual gas analyzer to be (90.93$\pm$0.05)\% {\xe} and (8.89$\pm$0.01)\% $^{134}$Xe; the amount of other xenon isotopes is negligible. 
The scintillation light due to energy deposits in the detector is monitored by 1325 17-inch and 554 20-inch photomultiplier tubes (PMTs) mounted on the 18-m-diameter spherical stainless-steel tank. The total photo-cathode coverage is 34\%. The stainless-steel tank is surrounded by the outer detector (OD) filled with 3.2\,kton of pure water. It shields the LS from external radiation and acts as Cherenkov veto for identifying cosmic ray muons using 225 20-inch PMTs. 

The data-acquisition system is triggered when the number of 17-inch PMTs hit is more than 70 within a 125\,nsec window, the so-called primary trigger. This corresponds to a threshold of $\sim$0.4\,MeV. After each primary trigger, the threshold is lowered to $\sim$0.25\,MeV for 1~ms to study sequential decays. The event energy (visible energy) and position are reconstructed based on the photon hit-time and charge distribution. The energy response is calibrated with three kinds of sources; (i) $^{208}$Tl 2.614\,MeV $\gamma$'s from an artificial ThO$_2$-W calibration source~\cite{KamLANDZen:2012aa}, (ii) 2.225\,MeV $\gamma$'s from spallation neutrons capturing on protons, and (iii) $^{214}$Bi ($\beta + \gamma$) from $^{222}$Rn ($\tau$ = 5.5\,days) introduced during the initial filling of the liquid scintillator. The energy resolution is estimated to be $\sigma=(6.6 \pm 0.3)\%/\sqrt{E({\rm MeV})}$ from the energy distribution of (i). The systematic variation of the energy reconstruction with position or time is monitored by (ii), and is less than 1.0\%. 
Energy nonlinearity effects caused by scintillator quenching and Cherenkov light production are constrained by the gamma peak of (i) and the spectral shape of (iii). The vertex resolution is estimated from the radial distribution of radioactive contaminants to be $\sigma\sim$15\,cm/$\sqrt{E({\rm MeV})}$~\cite{KamLANDZen:2012aa}. 

\begin{figure}[htbp]
	\begin{center}
	\includegraphics[angle=270,width=1\columnwidth,clip]{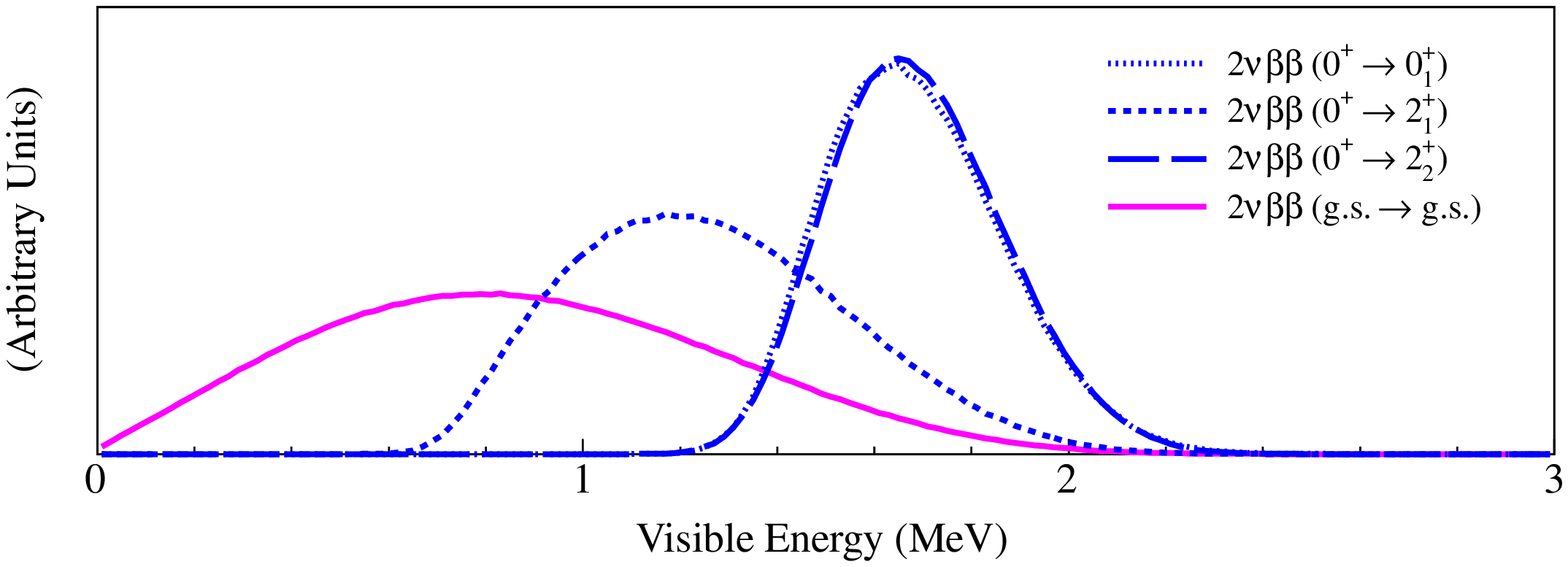}
	\includegraphics[angle=270,width=1\columnwidth,clip]{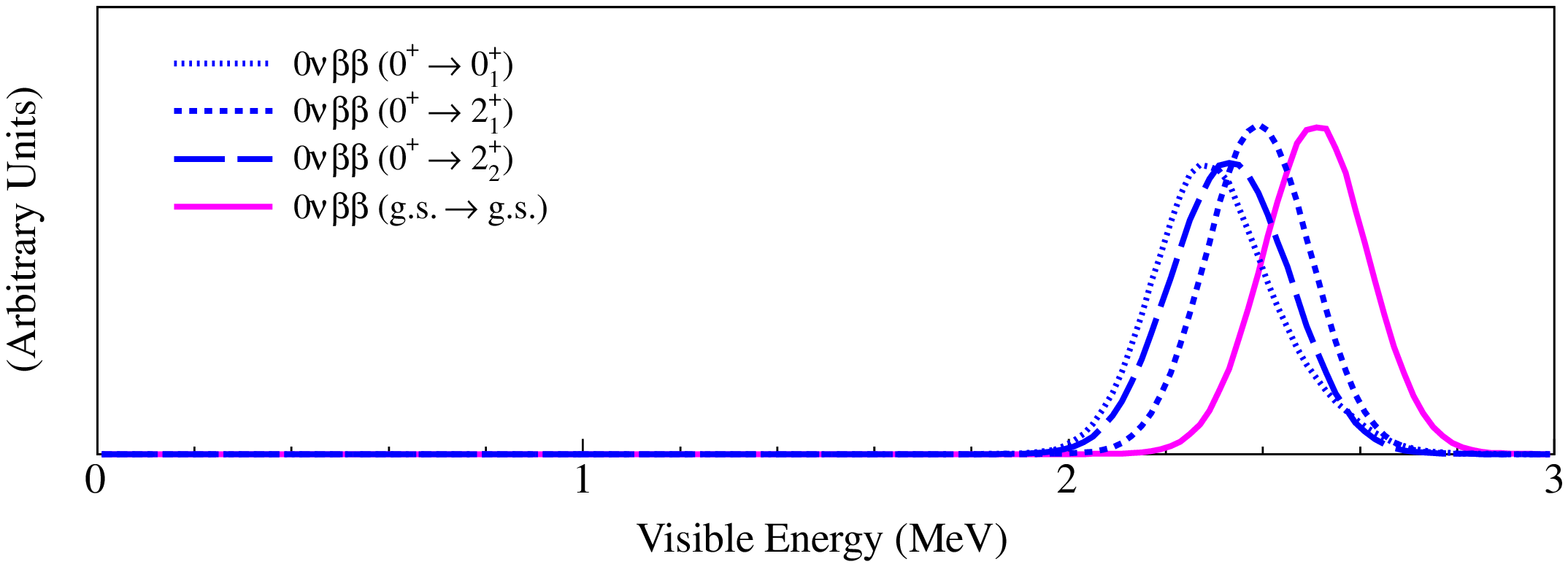}
	\end{center}
	\caption[]{Visible energy spectra of {\xe} double-beta decay to various excited states of $^{136}$Ba. The upper figure shows {\tnu} decays and the lower figure shows {\znu} decays. The {\tnu} and {\znu} decays to the ground state are also shown for reference. All decays are shown on a linear scale.} 
	\label{figure:model}
	\end{figure}

In \mbox{KamLAND-Zen}, the scintillation light of two coincident electrons generated from {\xe} double-beta decay cannot be separated, only their summed energy is observed. 
{\tnu} decay to the ground state has a continuous energy spectrum spanning from zero to the $Q$-value of {\xe}. Decays to the excited states have similar spectra, but they begin at different values of visible energy spectra depending on the energies of the de-excitation emissions.
In contrast to {\tnu} decay, {\znu} decay produces a mono-energetic peak at the $Q$-value with width determined by the energy resolution. The peak position is shifted for different decay modes since quenching effects of the liquid scintillator depend on the identities of the emitted particles (electron or $\gamma$ ray) and their energies. Fig.~\ref{figure:model} shows the predicted energy spectra of {\xe} double-beta decay to various excited states as well as to the ground state. The spectra have been convolved with the \mbox{KamLAND-Zen} detector response function, including energy resolution and particle-dependent energy scale non-linearities.

\section{Data sets and event selection }
\label{sec:event_selection}
The analysis is based on data collected from October 12, 2011 to June 14, 2012, the same period used in Ref.~\cite{Gando:2012zm}. This corresponds to the first phase data set. The data set is divided into two sets separated by detector maintenance work done in February 2012 aiming to remove radioactive impurities. In the earlier analysis~\cite{Gando:2012zm}, $^{110m}$Ag was identified as a likely background in the {\znu} decay search. During the maintenance, a  total of 37\,m$^3$, or 2.3 full volume exchanges, of Xe-LS were passed through a 50\,nm filter. The Xe-LS was circulated with two Teflon tubes placed at the bottom of a corrugated tube and the IB, which was inserted just before the filtration started and remained there until the second data taking period finished. During the filtration, $4.3 \pm 0.5$ kg of Xe is added to Xe-LS. The data-set period before the filtration is denoted as \mbox{DS-1} and the data-set period after filtration is denoted as \mbox{DS-2}. 

Double-beta decay candidate events are selected by performing the following series of cuts: (i) The reconstructed vertices of events are within the fiducial volume (FV) defined for each data set and transition. (ii) Muons and post-muon events within a time window of 2\,msec are rejected. Muons are identified by events which leave more than 10,000\,p.e. in the LS or produce more than 5 OD hits. (iii) Time and space correlated events which occurred within 3\,msec and 300\,cm are eliminated to avoid backgrounds from  sequential Bi-Po decays. The dead time introduced by this coincidence cut is less than 0.1\%.
(iv) A background mainly from reactor ${\overline{\nu}}_{e}$'s is eliminated by vetoing delayed coincidence of positrons (prompt signals) and neutron capture $\gamma$'s (delayed signals). (v) After applying the above selection criteria, any remaining noise events are rejected by a vertex-time-charge (VTQ) test. This test compares the observed charge and hit-time distribution with the expectation based on the reconstructed vertex, filtering unphysical events. 
This cut reduces the selection efficiency by less than 0.1\%. 
  
The {\tnu} and {\znu} analyses use different fiducial volume cuts to optimize the signal-to-background for the given energy region of interest. 
For {\tnu} decay, the main background is $^{134}$Cs (see Section~\ref{sec:bg}), predominantly found on the surface of the IB, so a reduction of the fiducial volume suppresses this background. On the other hand, the main backgrounds for the {\znu} decay analysis are $^{214}$Bi on the IB and $^{110m}$Ag in the Xe-LS. For \mbox{DS-1}, 1.2-m-radius and 1.35-m-radius cuts are found to be optimal for the {\tnu} and {\znu} decay analysis, respectively. For \mbox{DS-2}, the fiducial radii are the same as \mbox{DS-1}, but additional cuts were made to avoid backgrounds from Teflon tubes inserted into the IB between \mbox{DS-1} and \mbox{DS-2}. The additional cuts are a 0.2-m-radius cylindrical cut along the central vertical axis and a 1.2-m-radius spherical cut around the stainless steel inlet tip of the Teflon tube. The total livetime after removing periods of high background rate due to $^{222}$Rn daughters  introduced during filtration is 213.4\,days. The total {\xe} exposure is 61.8\,\mbox{kg$\cdot$yr} and 89.5\,\mbox{kg$\cdot$yr} for the search of {\tnu} and {\znu} decays to the excited states, respectively. The livetime, Xe concentration, fiducial Xe-LS mass, {\xe} mass, and {\xe} exposure for each data set are summarized in Table~\ref{table:fiducial}.
	\begin{table}[htbp]
	\begin{center}
	\caption{\label{table:fiducial}Data sets used for the analysis of {\xe} double-beta decay to excited states. R$_{1.20\rm{m}}$ (R$_{1.35\rm{m}}$) refers to the R $<$ 1.20 m (R $<$ 1.35 m) cut. R$_{1.20\rm{m}}^*$ (R$_{1.35\rm{m}}^*$) refers to the radius cut and the Teflon tube cuts that were introduced to avoid  potential backgrounds due to that tube. }
	\vspace{0.3cm}
	\begin{tabular}{@{}*{4}{lccc}}
	\hline
	\hline
	& ~~~\mbox{DS-1}~~~ & ~~~\mbox{DS-2}~~~ & ~~~Total~~~ \\
	\hline
	livetime (days) & 112.3 & 101.1 & 213.4 \\
	Xe concentration (\% by weight) & 2.44 & 2.48 & - \\
	\hline
	\multicolumn{1}{c}{{\tnu} decay} & R$_{1.20\rm{m}}$ & R$_{1.20\rm{m}}^*$ & \\
	\hline
	fiducial Xe-LS mass\,(ton) & 5.65 & 3.73 & - \\
	{\xe} mass\,(kg) & 125 & 84.0 & - \\
	{\xe} exposure\,(kg$\cdot$yr) & 38.6 & 23.2 & 61.8 \\
	\hline
	\multicolumn{1}{c}{{\znu} decay} & R$_{1.35\rm{m}}$ & R$_{1.35\rm{m}}^*$ & \\
	\hline
	fiducial Xe-LS mass\,(ton) & 8.04 & 5.55 & - \\
	{\xe} mass\,(kg) & 179 & 125 & - \\
	{\xe} exposure\,(kg$\cdot$yr) & 54.9 & 34.6 & 89.5 \\
	\hline
	\hline
	\end{tabular}
	\end{center}
	\vspace{-0.3cm}
	\end{table}

\section{Background estimation}
\label{sec:bg}
The main backgrounds to the double-beta decay study can be divided into three categories: (i) radioactive impurities from and external to the IB, (ii) those within the Xe-LS and (iii) spallation products generated by cosmic-ray muons. 
As reported in Ref.~\cite{KamLANDZen:2012aa}, the external background is dominated by $^{134}$Cs ($\beta + \gamma's$) in the energy region 1.2 $< E <$ 2.0 MeV (the {\tnu} energy window), most likely due to balloon film contamination by fallout from the Fukushima I reactor accident in March, 2011. The existence of $^{137}$Cs (0.662\,MeV $\gamma$) on the IB surface is also confirmed. The IB was fabricated 100 km from the Fukushima reactor site a few months after the accident. In the energy region 2.2 $< E <$ 3.0 MeV (the {\znu} energy window), the dominant external contaminant is $^{214}$Bi ($\beta + \gamma$'s) from the $^{238}$U chain. 

The main radioactivity in the Xe-LS comes from the U and Th decay chains. Their contaminations are investigated using $^{214}$Bi-$^{214}$Po ($\beta$-$\alpha$) and $^{212}$Bi-$^{212}$Po ($\beta$-$\alpha$) decays. Assuming secular equilibrium, the $^{238}$U and $^{232}$Th concentrations are estimated to be $(1.3 \pm 0.2) \times 10^{-16}$\,g/g and $(1.8 \pm 0.1) \times 10^{-15}$\,g/g, respectively~\cite{Gando:2012zm}. To allow for the possibility of decay chain non-equilibrium, however, the Bi-Po measurements are used to constrain only the rates for the $^{222}$Rn-$^{210}$Pb sub-chain of the $^{238}$U series and the $^{228}$Th-$^{208}$Pb sub-chain of the $^{232}$Th series, while other background rates in both series are left unconstrained. 

Neutrons produced by muons are captured on the main components of the liquid scintillator, protons~(2.225\,MeV $\gamma$) and $^{12}$C~(4.946\,MeV). In addition, neutrons may be captured on {\xe} (4.026\,MeV) or $^{134}$Xe (6.364\,MeV). The expected neutron capture fractions on protons, $^{12}$C, $^{136}$Xe and $^{134}$Xe are 0.994, 0.006, $9.5 \times 10^{-4}$, and $9.4 \times 10^{-5}$, respectively. The neutron capture product $^{137}$Xe ($\beta^{-}$, \mbox{$\tau=5.5$\,min}, \mbox{$Q=4.17 $\,MeV}) may be produced by {\xe} as a potential background for {\znu} decay, but its expected rate is negligible.
Event rates of carbon spallation products at \mbox{KamLAND-Zen} were obtained in Ref.~\cite{Abe:2009aa}. They are 
$1.11 \pm 0.18$\,\PerTonDay\, and $(2.11 \pm 0.18) \times 10^{-2}$\,\PerTonDay\ for $^{11}$C ($\beta^{+}$, \mbox{$\tau=29.4$\,min}, \mbox{$Q=1.98$\,MeV}) and $^{10}$C ($\beta^{+}$, \mbox{$\tau = 27.8$\,s}, \mbox{$Q = 3.65$\,MeV}), respectively.
For the muon induced products from xenon, we have no past experimental data. Spallation products from {\xe} whose half-live are less than 100\,sec, are studied with the present data. This is done by searching for time-correlations of events within a distance of 1.50\,m from track-reconstructed muons depositing more than $\sim$3 GeV of energy (so-called showering muons). Events in a time window of 0-500\,sec after muon passage are assumed to be potentially the decays of muon spallation products, while events in a time window of 500-5000\,sec are used to estimate the accidental background.
No correlated events are found in the energy range 1.2 $< E <$ 2.0 MeV and 2.2 $< E <$ 3.0 MeV. 

In addition to the three categories of background, a peak was found in the {\znu} energy window. As extensively discussed in Ref.~\cite{KamLANDZen:2012aa,Gando:2012zm}, we concluded that it is most likely due to a background from $^{110m}$Ag ($\beta^{-}$, \mbox{$\tau=360$\,days}, \mbox{$Q=3.01$\,MeV}) in Xe-LS. 
 Only four nuclei can give a possible background in this energy window: $^{60}$Co ($\beta^-$,\mbox{$\tau=7.61$\,yrs}, \mbox{$Q=2.82$\,MeV}), $^{88}$Y (electron capture (EC), \mbox{$\tau=154$\,days}, \mbox{$Q=3.62$\,MeV}), $^{110m}$Ag, and $^{208}$Bi (EC, \mbox{$\tau=5.31\times10^5$\,yrs}, \mbox{$Q=2.88$\,MeV}). Due to their different lifetimes and the increased exposure time of this data set, the event rate time variation in the energy range $2.2 < E < 3.0$ MeV exhibits a strong preference for the lifetime of $^{110m}$Ag~\cite{Gando:2012zm}.
In addition, the combined data set of \mbox{DS-1} and \mbox{DS-2} indicates that $^{110m}$Ag is also distributed on the IB. 
This activity is estimated from two-dimensional fits in radius and energy, assuming that the only contributions on the IB are from $^{214}$Bi and $^{110m}$Ag~\cite{Gando:2012zm}.
The fit results for the $^{214}$Bi and $^{110m}$Ag event rates on the IB are given in Table~\ref{table:BGrates} for the two data sets.
	\begin{table}[htbp]
	\caption{\label{table:BGrates}Fitted event rates of $^{214}$Bi and $^{110m}$Ag on the IB. }
	\begin{center}
	\begin{tabular}{@{}*{4}{lcc}}
	\hline
	\hline
	& ~~~\mbox{DS-1}~~~ & ~~~\mbox{DS-2}~~~ \\
	\hline
	$^{214}$Bi (\PerDay) & $19.0 \pm 1.8$ & $15.2 \pm 2.3$\\
	 $^{110m}$Ag (\PerDay) & $3.3 \pm 0.4$ & $2.2 \pm 0.4$ \\
	\hline
	\hline
	\end{tabular}
	\end{center}
	\end{table}
The $^{214}$Bi rates are estimated after removing $^{214}$Bi-$^{214}$Po sequential decays and are consistent between \mbox{DS-1} and \mbox{DS-2}, while the $^{110m}$Ag rates are consistent with the decay time of this isotope.
The rejection efficiencies ({\znu} window) of the FV cut $R < 1.35\,{\rm m}$ against $^{214}$Bi and $^{110m}$Ag on the IB are $(96.8 \pm 0.3)\%$ and $(93.8 \pm 0.7)\%$, respectively, where the uncertainties include the uncertainty in the absolute IB position. For $R < 1.20\,{\rm m}$, they are both $(99.2 \pm 0.1)\%$.

\section{Systematic uncertainties}
\label{sec:systematic}
Systematic uncertainties are summarized in Table~\ref{table:systematics} for the two data sets. 
The dominant contribution comes from the fiducial volume uncertainty. For DS-1, the Xe-LS volume in the 1.2-m-radius and 1.35-m-radius is 7.24 and 10.3\,m$^3$, respectively. The volume ratio of the fiducial volume to the total volume is \mbox{0.438 $\pm$ 0.005} and \mbox{0.624 $\pm$ 0.006}, where the total volume was measured by the flow meter to be~\mbox{(16.51 $\pm$ 0.17)\,m$^3$} during Xe-LS filling. 
The FV fraction is also estimated from the ratio of $^{214}$Bi events which pass the FV cuts to the total number in the entire Xe-LS volume after subtraction of the IB surface contribution. The results for DS-1 are $0.421 \pm 0.007({\rm stat}) \pm 0.001({\rm syst})$ and $0.620 \pm 0.007({\rm stat}) \pm 0.021({\rm syst})$ for 1.2-m-radius and 1.35-m-radius, respectively.
The difference in these values is taken as a measure of the systematic uncertainty on the vertex-defined FV. Combining the errors, we obtain 5.6\% and 3.9\% systematic uncertainties on the 1.20-m-radius and 1.35-m-radius for \mbox{DS-1}. Similarly, the error for \mbox{DS-2} is estimated to be 6.3\% and 4.1\%. The uncertainty for DS-1 at 1.20-m-radius is slightly different from Ref.~\cite{Gando:2012pj} due to improved energy/vertex reconstruction. 
The total systematic uncertainties for the double-beta decay half-life measurements for DS-1/DS-2 are 5.6\%/6.3\% and 3.9\%/4.1\%~\cite{Gando:2012zm}, obtained by adding all uncertainties listed in Table~\ref{table:systematics} in quadrature. 
	\begin{table}[htbp]
	\caption{\label{table:systematics}Estimated systematic uncertainties used for the {\xe} double-beta decay half-life measurement. }
	\begin{center}
	\begin{tabular}{@{}*{6}{lccccc}}
	\hline
	\hline
	Source  & \multicolumn{5}{c}{Systematic Uncertainty (\%)} \\
	\hline
	  & \multicolumn{2}{c}{~~~~~{\tnu} decay~~~~~} & \multicolumn{2}{c}{~~~~~{\znu} decay~~~~~}&Ref. \\
	\hline
	& ~~DS-1~~ & ~~DS-2~~ & ~~DS-1~~ & ~~DS-2~~  &  \\
	& R$_{1.20\rm{m}}$ & R$_{1.20\rm{m}}^*$ & R$_{1.35\rm{m}}$ & R$_{1.35\rm{m}}^*$ & \\
	\hline
	Fiducial volume & 5.6 & 6.3 & 3.9 & 4.1 & \cite{Gando:2012zm}$^\dagger$\\
	Enrichment factor of {\xe} & 0.05 & 0.05 & 0.05 & 0.05 & \cite{KamLANDZen:2012aa}\\
	Xenon concentration & 0.34 & 0.37 & 0.34 & 0.37 & \cite{Gando:2012zm}\\
	Detector energy scale & 0.3 & 0.3 & 0.3 & 0.3 & \cite{KamLANDZen:2012aa}\\
	Detection efficiency & 0.2 & 0.2 & 0.2 & 0.2 & \cite{Gando:2012zm}\\
	\hline
	Total & 5.6 & 6.3 & 3.9 & 4.1 & \\
	\hline
	\hline
	$^\dagger$ Only for {\znu} decay.
	\end{tabular}
	\end{center}
	\end{table}

\section{Results}
\begin{figure}[htbp]
	\begin{center}
	\includegraphics[angle=270,width=1.0\columnwidth]{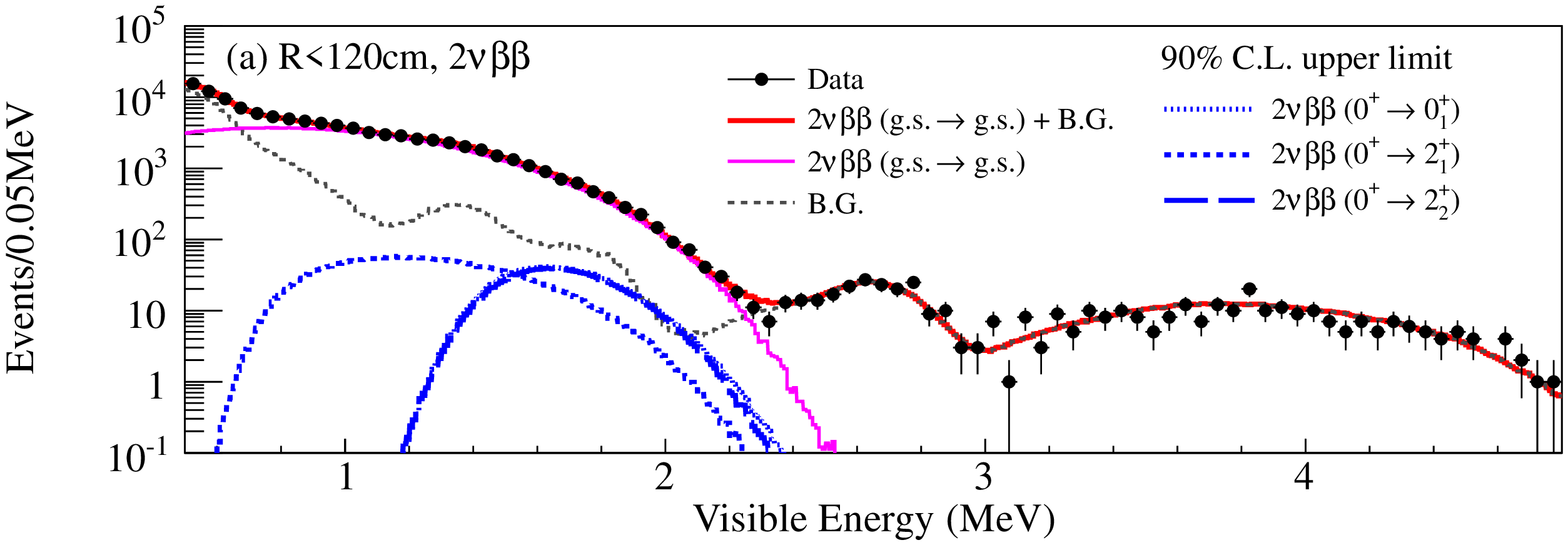}
	\includegraphics[angle=270,width=1.0\columnwidth]{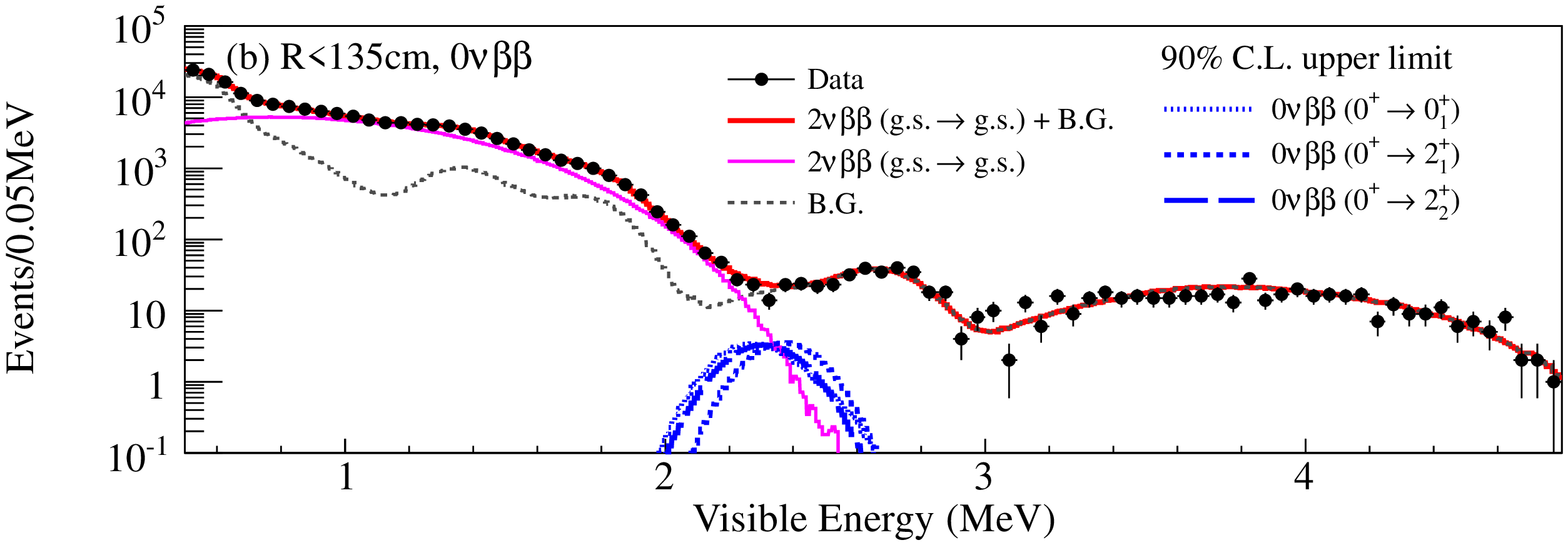}	
	\end{center}
	\caption[]{Energy spectrum of selected double-beta decay candidates (data points) with best-fit backgrounds (gray dashed curve) and total (red curve). Data of DS-1 and DS-2 are combined. The 90\% upper limits for {\xe} double-beta decay to the excited states of $^{136}$Ba are shown in blue for {\tnu} (top figure) and {\znu} (bottom figure). The peak around 0.5 MeV in the background spectrum mainly comes from $^{85}$Kr in the Xe-LS and from $^{137}$Cs in the IB. The two peaks at 1.2 $< E <$ 2.0 MeV are largely due to external $^{134}$Cs. The main contributions in the region 2.0 $< E <$ 3.0 MeV are $^{110m}$Ag in the Xe-LS and external  $^{214}$Bi from the $^{238}$U decay chain. In 3.0 $< E <$ 4.8 MeV range, the dominant backgrounds are  $^{208}$Tl ($^{232}$Th decay chain) in the Xe-LS and those from the IB.
	 }
	\label{figure:energy}
	\end{figure}

The upper limits of the {\xe} {\tnu} and {\znu}  decay rates to the excited states studied in the present work are estimated from a likelihood fit to the binned energy distribution between 0.5 and 4.8\,MeV for each data set except for {\tnu} decays to 0$^+_1$ and 2$^+_2$ excited states. The fit range of these two decays is between 1.2 and 2.3\,MeV. The {\tnu} decay spectra were calculated based on the approximated differential rate~\cite{Haxton1984} and Fermi function~\cite{Schenter1983}. Backgrounds from decay chains of $^{222}$Rn-$^{210}$Pb and $^{228}$Th-$^{208}$Pb in the Xe-LS, the spallation products $^{10}$C and $^{11}$C, contributions from the IB including the decay chains of $^{222}$Rn-$^{210}$Pb and $^{228}$Th-$^{208}$Pb, $^{134}$Cs, $^{137}$Cs and $^{110m}$Ag are allowed to vary in the fit, but are constrained by their estimated rates. 
In addition, the event rates of {\xe} {\znu}, $^{110m}$Ag, $^{208}$Bi, external $^{110m}$Ag and $^{222}$Rn-$^{210}$Pb are constrained by the time variation of the observed event rate in 2.2 $< E <$ 3.0 MeV considering their half-lives. The uncertainties of the energy scale parameters are constrained from the $^{208}$Tl calibration and radon-induced $^{214}$Bi data.
Other events, such as {\tnu} decays to the ground state and contributions from the Xe-LS such as $^{210}$Bi, $^{85}$Kr, and $^{110m}$Ag, are unconstrained in the fit. The energy spectra of selected events for the combined data from DS-1 and DS-2 are shown in Fig.~\ref{figure:energy}. 

To estimate the limits on decay to the 0$^+_1$, 2$^+_1$ and 2$^+_2$ excited states of $^{136}$Ba, each branch was individually fitted simultaneously with the dominant contribution from {\tnu} decay to the ground state. The 90\% C.L. upper limits on the number of  events in each transition are estimated from the single rate that gives $\Delta\chi^2$=2.71 compared with the best-fit spectrum for each data set (DS-1 and DS-2).
The resulting upper limits for each transition are drawn in Fig.~\ref{figure:energy}, and the corresponding half-life limits are summarized in Table~\ref{table:result}, together with theoretical estimates from Ref.~\cite{PhysRevC.91.054309}. 
	\begin{table}[htbp]
	\caption{\label{table:result} Half-life lower limits for {\xe} double-beta decay to excited states of $^{136}$Ba at 90\% C.L. Theoretical estimates of the half-life of the {\tnu} decays including the range of values calculated using the four models taken from Ref.~\cite{PhysRevC.91.054309}}
	\begin{center}
	\begin{tabular}{c c c c}
	\hline
	\hline
	\multirow{2}{*}{~~~Transition~~~}& \multicolumn{2}{c}{$T_{1/2}$ (yr, 90\% C.L.)}\\
	& ~~~This work~~~ & ~~~Previous work~~~&~~~Theoretical Est.~\cite{PhysRevC.91.054309}~~\\
	\hline
	{\tnu} decay & & \\
	$0^+ \rightarrow 0^+_1$ & $8.3\times10^{23}$ & - & $(1.3-8.9)\times10^{23}$\\
	$0^+ \rightarrow 2^+_1$ & $4.6\times10^{23}$ & $9.4\times10^{21}$~\cite{Bernabei:2002bn} & $(1.6-48)\times10^{25}$\\
	$0^+ \rightarrow 2^+_2$ & $9.0\times10^{23}$ & - & $(1.2-6.3)\times10^{30}$\\
	{\znu} decay & & \\
	$0^+ \rightarrow 0^+_1$ & $2.4\times10^{25}$ & - & \\
	$0^+ \rightarrow 2^+_1$ & $2.6\times10^{25}$ & $6.5\times10^{21}$~\cite{Zanotti:1991vh} & \\
	$0^+ \rightarrow 2^+_2$ & $2.6\times10^{25}$ & - & \\
	\hline
	\hline
	\end{tabular}
	\end{center}
	\end{table}
The limits of the {\tnu} and {\znu} decays to the $2^+_1$ excited state of $^{136}$Ba are improved by a factor 49 and 4.0$\times10^3$, respectively, relative to previous results in Refs.~\cite{Bernabei:2002bn,Zanotti:1991vh}. The limits on the {\tnu} and {\znu} decays to the $0^+_1$ and $2^+_2$ excited states are estimated for the first time.
The obtained limit for the {\tnu} decay to the $0^+_1$ excited state can constrain three of four theoretical models discussed in Ref.~\cite{PhysRevC.91.054309}. From the limit on the {\xe} {\znu} decay ($0^+ \rightarrow 0^+_1$) half-life, we obtain a 90\% C.L. upper limit of $\left<m_{\beta\beta}\right>$ $<$ 0.36-0.84 eV using nuclear matrix elements evaluated in Refs.~\cite{PhysRevC.64.035501,Suhonen201136}.

\section{Summary and conclusion}
We used the \mbox{KamLAND-Zen} first phase data set to search for {\tnu} and {\znu} decays of {\xe} to the 0$^+_1$, 2$^+_1$ and 2$^+_2$ excited states of $^{136}$Ba. No signal was observed and half-life limits at 90\% C.L. were obtained as summarized in Table~\ref{table:result}. The lower limits on $0^+ \rightarrow 0^+_1$ for {\tnu} and {\znu} decay were improved by a factor of 49 and 4.0$\times10^3$, respectively, from previous measurements.
The decays of $0^+ \rightarrow 0^+_1$ and $0^+ \rightarrow 2^+_2$ for {\tnu} and {\znu} decay were investigated for the first time. 
The limit of $0^+ \rightarrow 0^+_1$ for {\tnu} decay allows to constrain some theoretical predictions.

The current results for the corresponding {\znu} transitions are mainly limited by the background from $^{110m}$Ag and $^{214}$Bi. Distillation of the LS to remove the $^{110m}$Ag and replacement of the IB to reduce $^{214}$Bi should improve the sensitivity. 
The dominant background of the {\tnu} decay to the $0^+_1$ excited states comes from external $^{134}$Cs. Replacement of the IB will reduce this radioactivity and provide improved sensitivity to this decay.

\section{Acknowledgements}
The \mbox{KamLAND-Zen} experiment is supported by JSPS KAKENHI Grant Numbers 21000001 and 26104002; Stichting FOM in the Netherlands; and under the U.S. Department of Energy (DOE) Grant No.\,DE-AC02-05CH11231, as well as other DOE and NSF grants to individual institutions. The Kamioka Mining and Smelting Company has provided service for activities in the mine. We acknowledge the support of NII for SINET4.




\bibliographystyle{model1a-num-names}
\bibliography{DoubleBetaExcitedStates}







\end{document}